\documentclass[a4paper,12pt,twoside]{article}
\usepackage{psfig}
\usepackage{epsfig}
\usepackage{amssymb,wasysym,bbm,feynmf,rotating,subfigure,here,citesort}
\usepackage{graphicx}
\begin{document}

%
\pagestyle{empty}
%
%
\title{ \vspace*{0cm} {\normalsize\rightline{CU-TP-1151, BI-TP 2006/23}}
  \vspace*{0.cm} 
{\Large \bf
    \boldmath On Kolmogorov Wave Turbulence in QCD}\footnote{This work is supported in part by the US
  Department of Energy.}}

\author{}\date{} \maketitle
 
\vspace*{-2.5cm}
 
\begin{center}
 
\renewcommand{\thefootnote}{\alph{footnote}}
 
{\large
A.H. Mueller$^1$,
A.I. Shoshi$^2$ and
S.M.H. Wong$^1$ } 
 
 
{\it $^1$ Physics Department, Columbia University, New York, 
     NY 10027, USA \\
     $^2$ Fakult{\"a}t f{\"u}r Physik, Universit{\"a}t
     Bielefeld, D-33501 Bielefeld, Germany
}

\end{center}
 

\begin{abstract}

We investigate Kolmogorov wave turbulence in QCD or, in other words,
we calculate the spectrum of gluons as a function of time,
$f_k(t)$, in the presence of a source which feeds in energy density in
the infrared region at a constant rate. We find an early, an
intermediate and a late time form for the gluon spectrum. Wave
turbulence in QCD turns out to be somewhat different than the
turbulence in the case of $\phi^4$-type theories studied by Zakharov,
L'vov and Falkovich. The hope is that a good understanding of QCD wave
turbulence might lead to a better understanding of the instability
problem in the early stages of the evolution after a heavy ion
collision.

  \vspace{1.cm}
 
 

 
\end{abstract}

%
%
%
\pagenumbering{arabic}
\pagestyle{plain}
%

\section{Introduction}

One of the key questions in heavy ion physics is how rapidly and by
what mechanism equilibrium is reached after a collision.
Phenomenological analyses of experimental data at RHIC suggest a rapid
thermalization~\cite{Kolb:2001qz+X,Teaney:2003kp+X} but, so far, no
convincing theoretical picture has emerged which naturally gives such
a quick approach to equilibrium.

The ``bottom-up'' picture of equilibration~\cite{Baier:2000sb} is a
detailed picture of how the initially produced hard gluons (at or near
the saturation scale of the colliding ions) lose energy by radiating
softer gluons.  After they become sufficiently
numerous these softer gluons equilibrate amongst themselves and
continue to absorb energy from the initially produced hard gluons.
After the hard gluons have lost all their energy the system has
reached complete equilibration.  Although difficult to evaluate
precisely, the time for this equilibration was estimated to be on the
order of 3 fm~\cite{Baier:2002bt}.  While it is an attractive picture
the bottom-up scenario has a serious flaw in that the initially
produced hard particles quickly acquire an instability as they spread
out along the axis of the heavy ion collision as it was recently
worked out by Arnold, Lenaghan and Moore~\cite{Arnold:2003rq} and was
long ago advocated by Mrowczynski~\cite{Mrowczynski:1988dz+X} (For
other early disscusions, see
Refs.~\cite{Weibel:1959,Buneman:1958,Heinz:1985vf+X}).  This
instability then becomes the dominant mechanism for the creation of
softer gluons~\cite{Arnold:2003rq}, more important than the Bethe-Heitler radiation used in
the bottom-up picture~\cite{Baier:2000sb,Wong:1996va}.  The instability
potentially speeds up equilibration, but so far it has been difficult
to follow analytically how the QCD gluonic system evolves toward
equilibration in the presence of the
instability~\cite{Mueller:2005un+X,Bodeker:2005nv}.
Numerical
studies~\cite{Arnold:2005ef,Romatschke:2003ms+X,Rebhan:2004ur+X,Romatschke:2005pm+X,Dumitru:2005gp+X,Mrowczynski:2005ki}
seem to indicate that the instability is effective at early times, but
becomes less prominent in non-Abelian theories when the occupation
number of low momentum modes becomes large.  So far there is no
analytic understanding of the numerical results.

The phase space spectrum found in numerical studies by Arnold and
Moore~\cite{Arnold:2005ef} has a resemblance to that of the Kolmogorov
spectrum in turbulence~\cite{Arnold:2005qs}.  Indeed, the problem of
wave turbulence with an infrared source of energy discussed by
Zakharov, L'vov and Falkovich (ZLF)~\cite{ZLF} (for a nice overview,
see~\cite{Micha:2004bv}) would seem to have much in common with the
instability problem currently under discussion for a non-expanding QCD
medium and where the source for the instability is a collection of
hard particles having a fixed asymmetry in their momentum
distribution. Wave turbulence is a somewhat easier problem to deal
with since one can take a spherically symmetric source of inflow of
energy in low momentum modes.  One can hope that a good understanding
of QCD wave turbulence will lead to a better understanding of the
early stages of evolution after a high-energy heavy ion collision.

Wave turbulence in QCD appears to be somewhat different than the
situation studied by Zakharov, L'vov and Falkovich~\cite{ZLF}. The
essence of the ZLF discussion is that waves, or particles, interact
with each other locally in momentum.  This is certainly the case, say,
in a $\phi^4$ theory where the scattering cross section for a high
momentum particle on a low momentum particle is very small.  Soft
quanta in $\phi^4$ theory interact very weakly with hard quanta
because of the small overlap in their fields due to Lorentz
contraction of the harder quanta.  Soft gluons, on the other hand,
interact with the currents of the harder gluons, and these currents do
not decrease with the momentum of the harder gluons.  In addition a
gauge theory has a lower cutoff in frequency, the plasma frequencly,
which has no analog in a theory of the $\phi^4$ variety.  Because of
the strong interaction between soft and hard modes along with the
plasma frequency infrared cutoff QCD turbulence seems to be quite
different than the ZLF problem.

In the problems considered by ZLF a source, say, of energy inserts
energy at an infrared scale and a sink extracts that energy at a
higher scale.  The flow of energy then proceeds from a low momentum
scale, through intermediate energy scales and exits at some high
energy scale.  A closely related process omits the high energy sink
and has the energy flowing from low to high momentum through all
intermediate scales.  What is different in the QCD case is that energy
can be absorbed by very high energy particles from a low energy source
without the necessity of passing through intermediate scales.
Particle number, however, cannot be directly transferred from low to
high energy but must flow through intermediate scales, or be created
at high energy through inelastic reactions.  Thus in QCD there are
important, even dominant, interactions which are not local in
momentum.

In this paper we calculate the analytic form of the spectrum of gluons
in the presence of a source which feeds in energy density at a
constant rate $\dot{\epsilon}_0=m_0^5/\alpha .$ We suppose the energy
comes into our system uniformly in space in the form of gluons which
have a spherically symmetric momentum distribution and which are
inserted uniformly in phase space just above the plasma frequency
cutoff in a range $m < \omega <\bar{m}$ with $m$ the plasma frequency
and $\bar{m}$ on the order of $m.\ m_0$ is the single dimensionful
parameter in the discussion given in Sec.2 while in Sec. 3 we allow
the incoming energy to be spread uniformly in phase space in a region
$0 \le k \le k_0$ of momenta in which case $k_0$ is a separate
dimensional parameter if we choose $k_0$ to be a scale larger than and
independent of $m_0.$ We always suppose $\alpha,$ the gluonic
coupling, to be small.

At late times after the source has become active,
$m_0t>(1/\alpha)^{9/5}$ in case the source energy is deposited in
$m<\omega <\bar{m},$ the system of gluons is very close to thermal
equilibrium.  The incoming energy is transferred from the scale $m$ to
the scale given by the temperature $T$ by direct absorption of soft
gluons in an inelastic $3\to 2$ process and, parametrically equally as
important, by elastic scattering of soft gluons (scale $m$) on hard
gluons (scale $T$) with the soft gluons losing energy to the hard
gluons.  Both $m$ and $T$ are slowly increasing with time.

In the time domain $(1/\alpha)^{7/5}<m_0t < (1/\alpha)^{9/5}$ gluons
having momentum much greater than $m$ are in thermal equilibrium while
gluon occupation numbers are significantly above the thermal curve in
the domain $m<\omega < \bar{m}.$

Finally, in the early time domain, $1\ll m_0t<(1/\alpha)^{7/5},$ the
system is far from thermal equilibrium in both the high and low
momentum regimes.  If $p_0(t)$ is the maximum scale to which gluons
have evolved, then $f_{p_0}\gg 1$ while $f_k={c(t)\over \alpha}
{m\over \omega}$ in the domain $m \ll \omega \ll p_0.$ This latter
distribution is a sort of equilibrium distribution although it does
not match onto a genuine equilibrium distribution at the scale $p_0.$

While intermediate momentum scales obey $f_k\sim 1/k$ in all time
domains, so long as $m_0t\gg 1,$ there is, nevertheless, a flow of
energy and particle number from soft to hard scales through these
intermediate scales.  In the case of energy this flow is negligible
compared to the direct transfer of energy from soft to hard scales,
while in the case of particle number flow it is one of the dominant
mechanisms of gluon number growth at the hard scale.

The dynamics which we use throughout this paper is the Boltzmann equation with a
collision term consisting of $2 \leftrightarrow 2$ and $2 \leftrightarrow 3$
gluon processes. We have assumed the absence of important longrange coherent
fields which, if present, could create problems for our approach.

\section{The spectrum of gluons as a function of time}

The problem we are concerned with here is very similar to the problem
of wave turbulence discussed by Zakharov, L'vov and Falkovich~\cite{ZLF}.
However, Zakharov et al. consider scalar theories where interactions
among particles are relatively local in momentum while the case of
interest here, QCD, allows important interaction between low momentum
and high momentum particles.  In addition in a gluonic theory there is
a minimum allowed frequency, the plasma frequency, at which waves can
propagate while a corresponding phenomenon does not exist with purely
scalar interactions.

At time, $t,$ equal to zero we turn on a source of gluons which feeds
a rate of increase of energy density $\dot{\epsilon}_0={dE\over d^3x
dt}={m_0^5\over \alpha}$ which is constant in time and uniform in
space.  The energy is incoming isotropically in low momentum modes but
as the density of gluons increases in time the source is modified so
that the incoming gluons always have energy just above the plasma
frequency, say $m < \omega < \bar{m}$ where $m$ is the plasma
frequency and $\bar{m}$ is of the same order as $m.$ The question we
wish to answer concerns the occupaton number, $f_{\vec{k}}$, of gluons
as a function of time.  As we shall see, the form of the spectrum
changes over the course of time having an early time, an intermediate
time and a late time form.  The late time form of $f_{\vec{k}}$ is the
simplest and the most straightforward to describe so we begin there.
Throughout we limit our discussion to the case of a fixed QCD
coupling, $\alpha$.

\subsection{The spectrum for $m_0t>(1/\alpha)^{9/5}$}

At sufficently large time it is apparent that the system will be
extremely close to thermal equilibrium since the source can only be a
very small perturbation on a system having a lot of energy. (The
scaling solution for constant energy flow found by ZLF~\cite{ZLF} (see also~\cite{Micha:2004bv}),
$f_k \sim (m/k)^{-5/3}$, connects an infrared source of energy to a
quantum equilibrium solution of the Boltzmann equation which exists
for $k>k_0$ whwere $f_{k_0}=1$. The non equilibrium solution in the
intermediate region of $k$ is necessary in a $\phi^4$-type theory
considered by ZLF where energy flow is local in
momentum. In the QCD case energy is transferred directly from the low
momentum source to the high momentum modes allowing the intermediate
momenta also to be in equilibrium. Indeed, we shall see throughout
this paper that $f_k \sim m/k$ is required, in intemediate momentum
regions, in order that the rate of emission and absorbtion of gluons
of momentum $k$ by harder gluons exactly cancel.) Thus energy
conservation gives
\begin{equation}
\epsilon (t) = {m_0^5\over \alpha} \cdot t=g_E T^4
\label{eq:eps_inflow}
\end{equation}
with $g_E=2(N_c^2-1){\pi^2\over 30},$ or
\begin{equation}
T = m_0 \left({m_0t\over g_E\alpha}\right)^{1/4}.
\label{eq:T_eq}
\end{equation}
The corresponding plasma frequency is
\begin{equation}
\omega_P \equiv m = {\sqrt{{4\pi\over 9}\alpha N_c}}\  T.
\label{eq:freq}
\end{equation}
and is related to the Debye mass by $m = m_D/\sqrt{2}$. At large times
the energy source inserts gluons into the system just above an energy
$m$ which is far below the temperature of the system $T.$ What we now
wish to describe is how the energy gets transferred from the scale $m$
to the scale $T$ which is where almost all the energy is located.
First we remark that, for very small $\alpha,$ the spectrum is given
by
\begin{equation}
f_k = {1\over e^{\omega/T}-1}
\label{eq:f_k_eq}
\end{equation}
which, when $\omega/T\ll 1,$ can be approximated as
\begin{equation}
f_k\simeq {T\over \omega}.
\label{eq:f_k_app_1/k}
\end{equation}
When $k/m\gg 1$ we shall not distinguish between $\omega$ and $k$ in
which case (\ref{eq:f_k_app_1/k}) could as well be written as
\begin{equation}
f_k\simeq {T\over k}.
\label{eq:f_k_app_1/k_wlk}
\end{equation}

Of course if the system were in exact equilibrium there would be no
transfer of energy and (\ref{eq:f_k_app_1/k_wlk}) would represent a system in complete
equilibrium without flow of energy or particle number.  Because we
have a source of incoming energy the occupation number will be
slightly higher than the equilibrium solution in the low momentum
region.  As we shall now demonstrate these low momentum gluons
directly transfer energy to the hard gluons, having momentum $T,$ with
only a small fraction of energy flowing from the scale $m$ through an
intermediate scale $k, m\ll k \ll T,$ and then to the scale $T.$
However, particle number, which must increase as $T^3,$ cannot be
transferred directly from low to high momentum but must pass through
intermediate scales giving a flow of particles from infrared to
ultraviolet in our spectrum.
\begin{figure}[htb]
\setlength{\unitlength}{1.cm}
\begin{center}
\epsfig{file=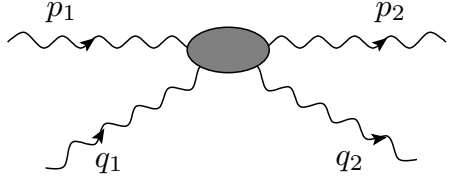, width=6cm}
\end{center}
\caption{Elastic gluon interactions.}
\label{elas_scatt}
\end{figure}

Consider the elastic scattering of soft gluons, $q_1$ and $q_2$, with
hard gluons having $p_1$, $p_2$ on the order of $T$. The process is
illustrated in Fig.~\ref{elas_scatt} and gives a flow of energy from
soft to hard as
\begin{eqnarray}
\dot{\epsilon}^{el} &=& 
{[2(N_c^2-1)]^2\over (2\pi)^{12}}\int_R{d^3p_1\over 2E_1}{d^3p_2\over 2E_2}
{d^3q_1\over 2\omega_1}{d^3q_2\over 2\omega_2}(2\pi)^4\delta^4(p_1+q_1-p_2-q_2)
\cdot \vert M\vert^2\cdot \nonumber \\
&& \times 
\left[f_{q_1}f_{p_1}(1+f_{q_2})(1+f_{p_2})-
f_{q_2}f_{p_2}(1+f_{q_1})(1+f_{p_1})\right](\omega_1-\omega_2)
\label{eq:eps_el_1}
\end{eqnarray}
where 
\begin{equation}
\vert M\vert^2 = \frac{64\pi^4}{N_c^2-1}\left({\alpha N_c\over \pi}\right)^2\left[3-{ut\over s^2}-{us\over t^2}-{ts\over u^2}\right]
\label{eq:M}
\end{equation}
with 
\begin{equation}
s=(p_1+q_1)^2, t=(q_1-q_2)^2, u=(p_1-q_1)^2.
\label{eq:mandeltam}
\end{equation}
The symbol $R$ in the integration in (\ref{eq:eps_el_1}) restricts
$\omega_{q_1}$, $\omega_{q_2}$ to be less than $\omega_{p_1}$,
$\omega_{p_2}$.
Now the hard particles $p_1$ and $p_2$ will have a distribution given
by (\ref{eq:f_k_eq}) so
\begin{equation}
1+f_p=f_p e^{E/T}
\label{eq:fp_eq}
\end{equation}
allows one to write the bracket in (\ref{eq:eps_el_1}) as
\begin{equation}
[\ ] = e^{E_{1/T}} f_{p_1} f_{p_2}f_{q_1}f_{q_2} 
\left[(\omega_1-\omega_2)/T+1/f_{q_2}\ -1/f_{q_1}\right].
\label{eq:brac_equil}
\end{equation}
If $f_{q_1}$ and $f_{q_2}$ also took the form (\ref{eq:f_k_eq}), or (\ref{eq:f_k_app_1/k}), we would
get exactly zero as expected in exact equilibrium.  However, the
incoming flux of soft gluons will naturally increase $f_q$ a little so
that $[\ ](\omega_1-\omega_2)$ is greater than zero corresponding to a
flow of energy from soft to hard particles. We can estimate that flow
of energy by rewriting (\ref{eq:eps_el_1}) as
\begin{equation}
\dot{\epsilon}^{el}=
{[2(N_c^2-1)]^2\over (2\pi)^8)}\int_R{d^3p_1d^3q_1\over 16 E_1E_2\omega_1\omega_2} 
d\Omega_{q_2}{q_2^2\over d\omega_2/dq_2}\vert M\vert^2(\omega_1-\omega_2)[\ ].
\label{eq:eps_el_eval}
\end{equation} 
Taking $\omega_1-\omega_2$ to be positive (Eq.~(12) is even under
$q_1\leftrightarrow q_2$) and
\begin{equation}
\vert M\vert^2 \sim \alpha^2 {(E\omega)^2\over m^4}
\label{eq:M_el_app}
\end{equation}
\begin{equation}
d^3p_1 \sim T^3
\label{eq:dp3_to_T3}
\end{equation}
\begin{equation}
d^3q_1 \sim  m^3
\end{equation}
\begin{equation}
q_2^2 \sim m^2
\end{equation}
\begin{equation}
\omega_1-\omega_2 \sim m
\end{equation}
along with (\ref{eq:f_k_eq}) and (\ref{eq:f_k_app_1/k}) one gets
\begin{equation}
\dot{\epsilon}^{el} \sim 
{m_0^5\over \alpha}\left({m\over m_0}\right)^5 
(\alpha f_{q_1})(\alpha f_{q_2}) 
\left[(\omega_1-\omega_2) + T/f_{q_2}\  - T/f_{q_1}\right]{1\over m}.
\label{eq:eps_el_app_param}
\end{equation}
Using
\begin{equation}
m/m_0 \sim (\alpha m_0t)^{1/4}
\label{eq:m_m0_eq}
\end{equation}
from (\ref{eq:T_eq}) and (\ref{eq:freq}) and
\begin{equation}
\alpha f_q \sim \alpha T/m \sim {\sqrt{\alpha}}
\label{eq:f_q_eq}
\end{equation}
from (\ref{eq:T_eq}) and (\ref{eq:f_k_app_1/k}) one finds
\begin{equation}
\dot{\epsilon}^{el} \sim {m_0^5\over \alpha} \alpha^{9/4}(m_0t)^{5/4} 
\left[(\omega_1-\omega_2) + T/f_{q_2} - T/f_{q_1}\right]{1\over m} \ .
\label{eq:eps_el_app_fin}
\end{equation}
Thus so long as $m_0t\gg \alpha^{-9/5}, f_q$ will be very close to the
equilibrium distribution making the [\ ] in (\ref{eq:eps_el_app_fin}) small and
compensating $\alpha^{9/4}(m_0t)^{5/4}.$

Although energy is transferred mostly in a direct way from the
incoming particles of the source to particles having $p\sim T$ there
is also interaction with particles having momentum $k$ where $m\ll k
\ll T.$ We can evaluate the rate of transfer of energy from (\ref{eq:brac_equil}) and
(\ref{eq:eps_el_eval}) by the replacement $p_1, p_2 \rightarrow k_1, k_2$ along with the
replacement $d^3p_1\rightarrow d^3k_1\sim k^3$ in (\ref{eq:dp3_to_T3}).  One easily
finds
\begin{equation}
\dot{\epsilon}^{el}_k\sim\left({m_0^5\over \alpha m}\right)\left({m\over m_0}\right)^5(\alpha f_{q_1})(\alpha f_{q_2})\left[\omega_1-\omega_2 + T/f_{q_2}-T/f_{q_1}\right]k/T
\label{eq:eps_k_T}
\end{equation}
after using (\ref{eq:f_k_app_1/k_wlk}).  $\dot{\epsilon}_k$ is the flow of energy from the
particles coming from the source to those particles having momentum on
the order of $k.$ As we have seen earlier with the convention
$\omega_1>\omega_2,$ [\ ] is positive so there is a positive flow of
energy from the source to $k$-particles, but this flow is suppressed
by a factor $k/T$ compared to the flow to particles of momentum $T.$
However, since $f_k$ changes only very slowly this flow of energy must
correspond to a flow of particles toward higher momentum.  We can
characterize this flow by a parameter, $\dot{k}_{flow},$ giving the
average change in a particle's momentum with time in terms of which
\begin{equation}
\dot{\epsilon}^{el}_k=2{(N_c^2-1)\over (2\pi)^3}f_k 4\pi  k^3 \dot{k}_{flow}.
\label{eq:eps_k_fl}
\end{equation}
In turn we can view $\dot{\epsilon}^{el}_k$ as the flow of particles past a
given momentum $k$ or
\begin{equation}
\dot{\epsilon}^{el}_k=k\dot{\rho}_k
\label{eq:eps_p_fl}
\end{equation}
where $\dot{\rho}_k$ is the number of particles per unit volume which
pass momentum $k,$ from lower toward higher momenta, in a unit of
time
\begin{equation}
\dot{\rho}_k=2{(N_c^2-1)\over (2\pi)^3} f_k 4\pi k^2\dot{k}_{flow}.
\label{eq:rho_fl}
\end{equation}
Using (\ref{eq:eps_p_fl}) and (\ref{eq:eps_k_T}) one sees that
$\dot{\rho}_k$ is independent of $k$ so that there is a constant, in
$k,$ flow of particles from lower to higher momenta.  One can check,
using (\ref{eq:eps_k_T}) and (\ref{eq:eps_p_fl}) along with (\ref{eq:rho_fl}) and $[(\omega_1-\omega_2) +
T/f_{q_2} - T/f_{q_1}]= m/(\alpha^{9/4}(m_0t)^{5/4})$ from (\ref{eq:eps_el_app_fin}), that
$\dot{\rho}_k\sim dT^3/dt$ so that the flow of particles is exactly
the right amount to furnish the increase of the total number of
particles of the system.
\begin{figure}[ht]
\setlength{\unitlength}{1.cm}
\begin{center}
\epsfig{file=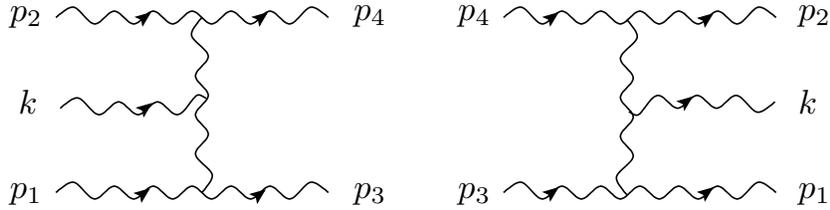, width=11cm}
\end{center}
\caption{$3 \leftrightarrow 2$ gluon processes.}
\label{inelas__23_32_scatt}
\end{figure}

While the number of gluons in the system is growing with time
$\dot{\rho}_k$ is much less than the rate at which gluons enter the
system, ${m_0^5\over \alpha m}.$ Thus inelastic reactions are
necessary in order to have chemical equilibrium.  The dominant
inelastic processes are those illustrated in Fig.~\ref{inelas__23_32_scatt} where gluons
1,2,3,4 have a momentum on the order of $T$ and $k\ll T.$
Dimensionally these graphs give (see Appendix \ref{App:2_3_in})
\begin{equation}
\dot{\epsilon}^{inel}\sim{T^6\omega\alpha^3\over m^2} 
\left[-f_{p_3}f_{p_4}(1+f_{p_1})(1+f_{p_2})(1+f_k) + 
f_{p_1}f_{p_2}f_k(1+f_{p_3})(1+f_{p_4})\right]
\label{eq:2_3_in}
\end{equation}
or, using (\ref{eq:T_eq}), (\ref{eq:freq}) and (\ref{eq:f_k_eq})
\begin{equation}
\dot{\epsilon}^{inel} \sim {m_0^5\over \alpha} 
\alpha^{9/4}(m_0t)^{5/4}\left[\omega - T/f_k\right]\cdot {1\over m}
\end{equation}
which is, parametrically, the same as (\ref{eq:eps_el_app_fin}).  Again, when $m<\omega
<\bar{m}$ one needs $f_k$ somewhat greater than $T/\omega$ in order to
transfer energy and lower the particle density to compensate for the
gluons emerging from the source. We note that the LPM effect is small
in the present case since the formation time of a gluon of momentum
$k$ is
\begin{equation}
\tau_f(k)\sim {2k\over k_t^2} \sim {2k\over m^2\cdot \tau_f/\lambda}
\end{equation}
or
\begin{equation}
\left({\tau_f\over \lambda}\right)^2\sim {k\over m^2\lambda} \sim {k\over T} < 1
\end{equation}
with $\lambda$ the mean free path for the particle to undergo a
collision of momentum transfer $m,$
\begin{equation}
{1\over \lambda} \sim {T^3\alpha^2\over m^2}.
\end{equation}

Finally we evaluate the mean free path, $\lambda_k$, for a particle of
momentum $k$ to undergo a collision of momentum transfer $k.$ Then,
using (\ref{eq:T_eq}),
\begin{equation}
{\lambda_k\over t} \sim \frac{1}{t k^3 f^2_k \alpha^2/k^2} \sim \left({k\over T}\right)\left[\alpha^7(m_0t)^5\right]^{-1/4},
\label{eq:lambda_mfp}
\end{equation}
so that we can expect the equilibrium spectrum $f_k\simeq T/k$ for $
\omega > \bar{m}$ so long as $m_0t > \alpha^{-7/5}.$ The case $m <
\omega < \bar{m}$ will be discussed in the next section.
%
\subsection{ The spectrum for $(1/\alpha)^{7/5} < m_0t < (1/\alpha)^{9/5}$}

\begin{figure}[htb]
\setlength{\unitlength}{1.cm}
\begin{center}
\epsfig{file=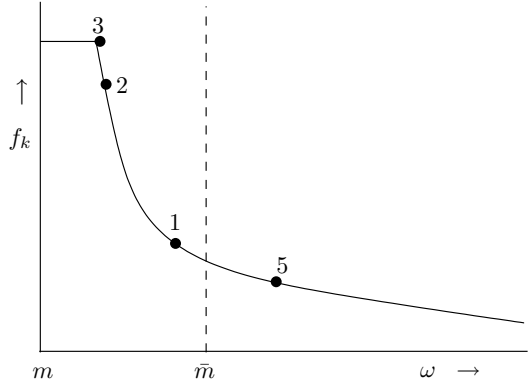, width=7cm}
\end{center}
\caption{Gluon distribution as a function of frequency.}
\label{gluon_dis}
\end{figure}

A major change occurs when $m_0t=(1/\alpha)^{9/5}.$ We have seen that
when $m_0t > (1/\alpha)^{9/5}$ the gluon system has a equilibrium
solution although there is a constant, in momentum, flow of particles
from the infrared region to the ultraviolet region.  However, from
(\ref{eq:eps_el_app_fin}) it is clear that when $m_0t < (1/\alpha)^{9/5}$ it is not
possible to transfer the energy from the source in the region $m <
\omega < \bar{m}$ to the momentum region near $T$ fast enough using a
near equilibrium distribution in the soft region.  What happens is
physically clear: The gluons from the source ``pile up'' in the region
$m < \omega < \bar{m}$ until the occupation number becomes large
enough to speed up the rate of transfer of energy from soft to hard
modes so that the transfer can exactly compensate the rate of energy
coming in from the source.  For $\omega > \bar{m}$ in the distribution
of gluons will be very close to the equilibrium distribution (\ref{eq:f_k_eq}),
however, when $m < \omega\, < \bar{m}$ the distribution will be
significantly altered.  We now turn to a description of this altered
distribution, illustrated in Fig.~\ref{gluon_dis}.

We suppose that the source creates gluons uniformly in phase space in
the region $m < \omega < \bar{m},$ with a constant inflow of energy
$\dot{\epsilon}=m_0^5/\alpha$ defining the parameter $m_0.$ This
corresponds to an increase in occupation number $\dot{f}_k\sim {m\over
\alpha}({m_0\over m})^5.$ Gluons having higher momentum, but located
in the region $m < \omega < \bar{m}$ will elastically scatter with
gluons having momentum on the order of $T$ and will lose energy to the
harder gluons as given by (\ref{eq:eps_el_app_param}).  Thus a gluon
at point 1 in Fig.~\ref{gluon_dis} will move to point 3 further increasing $f_{k_3}$
while $f_{k_1}$ will be determined by the competition between the
incoming gluons causing $f_{k_1}$ to increase and the scattering with
hard gluons causing $f_{k_1}$ to decrease.  Thus $f_k$ will increase
rapidly as $k$ decreases until $f_k$ becomes large enough that gluons
having momentum $k$ are absorbed by hard gluons, according to the
graphs in Fig.~\ref{inelas__23_32_scatt}, at the same rate that are
created from the external source and from gluons which arrive at $k$
after having elastically scattered with hard gluons.  We now give
estimates of these rates.

Suppose $k_3$ (point 3 in Fig.~\ref{gluon_dis}) is the momentum at which the
loss by inelastic absorption, the graphs in Fig.~\ref{inelas__23_32_scatt}, exactly balances
the rate of arrival of gluons coming directly from the source or
coming from the source via higher momentum regions.  Then the rate of
arrival of gluons is
\begin{equation}
\dot{f}_{k_3}^{source} \sim 
{m\over \alpha}\left({m_0\over m}\right)^5\left({m\over k_3}\right)^3
\label{eq:f_source}
\end{equation}
where ${m_0^5\over \alpha} \cdot {1\over m}$ is the rate of the total
number of gluons arriving over the whole phase space $m < \omega <
\bar{m}$ and $\left({m\over k_3}\right)^3$ expresses the fact that
these gluons end up in the limited region of phase space on the order
of $k_3^3.$ The rate at which gluons are absorbed by hard gluons,
according to the graphs in Fig.~\ref{inelas__23_32_scatt} is
\begin{equation}
\dot{f}_{k_3}^{abs} \sim - {T^6\alpha^3\over m^5} 
\left[f_{p_1}f_{p_2}f_{k_3}(1+f_{p_3})(1+f_{p_4})-
f_{p_3}f_{p_4}(1+f_{p_1})(1+f_{p_2})(1+f_{k_3})\right]
\end{equation}
or, using thermal distributions for $f_{p_1}, f_{p_2}, f_{p_3},
f_{p_4}$
\begin{equation}
\dot{f}^{abs}_{k_3} \sim - m{\sqrt{\alpha}}\ f_{k_3}\  
\left[{\omega\over T} - 1/f_{k_3}\right] T/m.
\label{eq:f_abs_k3}
\end{equation}
Dropping the $1/f_{k_3}$ term in (\ref{eq:f_abs_k3}) gives
\begin{equation}
\dot{f}^{abs}_{k_3}\sim - m {\sqrt{\alpha}} f_{k_3}
\label{eq:f_abs_k3_app}
\end{equation}
where we have anticipated $k_3/m \ll 1$ giving $\omega \simeq m.$

Now we must also require that the rate of $k_3$-gluons cascading to
even smaller momenta is not large compared to the rates given by (\ref{eq:f_source})
and (\ref{eq:f_abs_k3_app}).  Suppose $a\, k_3$-gluon goes to a momentum $k_4$ via
elastic scattering as given by Fig.~\ref{elas_scatt}, with the replacement $q_1 \to
k_3, q_2 \to k_4.$ Then (see Appendix \ref{App:3_4_el})
\begin{equation}
\dot{f}_{k_3}^{3 \to 4}\sim - T^3 {\alpha^2\over m^2}
\left({k_4\over m}\right)^3{m\over k_3}\\
\left[f_{p_1}f_{k_3}(1+f_{p_2})(1+f_{k_4})-f_{p_2}f_{k_4}(1+f_{p_1})(1+f_{k_3})\right]
\label{eq:3_4_el}
\end{equation}
which gives
\begin{equation}
\dot{f}_{k_3}^{3\to 4}\sim - m {\sqrt{\alpha}} f_{k_3} {\sqrt{\alpha}} 
f_{k_4}\left({k_4\over m}\right)^3{m\over k_3} \left[{\omega_3-\omega_4\over T} + 
1/f_{k_4}- 1/f_{k_3}\right] {T\over m}.
\end{equation}
Dropping the $1/f_{k_3}$ and $1/f_{k_4}$ terms and using
$\omega_3-\omega_4 \sim {1\over m} (k_3^2-k_4^2) \sim {k_3^2\over m}$,
since $k_4 \leq k_3 \ll m$, one gets
\begin{equation}
\dot{f}_{k_3}^{3\rightarrow 4} \sim - m {\sqrt{\alpha}} f_{k_3}{\sqrt{\alpha}} 
f_{k_4} \left({k_4\over m}\right)^3 \left({k_3\over m}\right) \ .
\label{eq:f_k3_3_to_4}
\end{equation}
Since we are looking for the value of $k_3$ at which the cascading to
lower momenta ceases it is natural to require that
(\ref{eq:f_source}), (\ref{eq:f_abs_k3_app}) and
(\ref{eq:f_k3_3_to_4}) be of the same size when $k_4$ is of the same
size as $k_3$.  This gives, from (\ref{eq:m_m0_eq}),
(\ref{eq:f_source}), (\ref{eq:f_abs_k3_app}) and
(\ref{eq:f_k3_3_to_4})
\begin{equation}
k_3/m \sim \left[\alpha^{9/5} m_0t\right]^{5/4}
\label{eq:k3_is}
\end{equation}
\begin{equation}
{\sqrt{\alpha}} f_{k_3} \sim \left[\alpha^{9/5} m_0t\right]^{-5} \ .
\label{eq:fk3_is}
\end{equation}

Finally, we give a calculation which shows that although $f_k$
increases sharply with decreasing $k,$ in the region $m < \omega <
\bar{m},$ it does not undergo a large change as $\omega$ goes from
above $\bar{m}$ to below $\bar{m}.$ To that end we first argue that in
the region $\omega > \bar{m}$ the occupancy $f_k$ lies not far from
the thermal curve.  Consider transitions $k_5\to k_3$ by elastic
scattering where $\omega_5$ is just above $\bar{m}$.  Then, neglecting
$1/f_{k_3}$ compared to $1/f_{k_5}$ (see Appendix \ref{App:3_4_el}),
\begin{equation}
\dot{f}_{k_5}^{5\rightarrow 3}\sim - m {\sqrt{\alpha}} f_{k_5}\left[{\sqrt{\alpha}} f_{k_3}\left({k_3\over m}\right)^3\right]
\left[{\omega_5-m\over T}-1/f_{k_5}\right] T/m.
\label{eq:fk5_5_3}
\end{equation}
The term ${\sqrt{\alpha}} f_{k_3}({k_3\over m})^3$ is very large so
that $\dot{f}_{k_5}^{5\to 3}$ can be small only if
\begin{equation}
f_{k_5} \simeq {T\over \omega_5-m}
\label{eq:fk5_is}
\end{equation}
which is close to the thermal curve. Now consider transitions, due to elastic scattering on hard gluons, where $k_1$ goes to $k_3.$  One has (see Appendix \ref{App:3_4_el})
\begin{equation}
\dot{f}_{k_3}^{1\rightarrow 3}\sim m{\sqrt{\alpha}} f_{k_1}{\sqrt{\alpha}} f_{k_3}\left[{\omega_1-m\over T}-1/f_{k_1}\right] T/m.
\label{eq:fk3_1_3}
\end{equation}
But the $\dot{f}^{1\rightarrow 3}_{k_3}$ in (\ref{eq:fk3_1_3}) cannot dominate (\ref{eq:f_abs_k3_app})
which means ${\sqrt{\alpha}} f_{k_1}$ cannot be large.  Since (\ref{eq:fk5_is})
means ${\sqrt{\alpha}} f_{k_5}$ is of order one, we see that there can
be no strong change in $f_k$ as $\omega$ crosses the value $\bar{m}$,
despite the fact that the source turns on abruptly for $\omega <
\bar{m}.$

\subsection{The spectrum for $m_0t <(1/\alpha)^{7/5}.$}

 From (\ref{eq:lambda_mfp}) one sees that the mean free path divided by time for hard
 particles becomes less than one when $m_0t < (1/\alpha)^{7/5}$ and
 thus that the majority of the gluons, and most of the energy, of the
 system are not in thermal equlibrium in the early time domain.
 Suppose, at a time $t,$ the gluon spectrum has reached momentum
 $p_0(t).$ Then energy conservation gives
\begin{equation}
{m_0^5\over \alpha} t \sim f_{p_0} p_0^4
\end{equation}
or, using
\begin{equation}
m^2 \sim \alpha f_{p_0} p_0^2,
\label{eq:m_fst}
\end{equation}
one gets
\begin{equation}
p_0 m \sim m_0^2{\sqrt{m_0t}}.
\label{eq:p0_fst}
\end{equation}
For $k/m \gg 1$ one expects $f_k$ to have the form
\begin{equation}
f_k \simeq \frac{c(t) m}{\alpha \omega} \ .
\label{eq:f_k_1/k}
\end{equation}
The $1/k$ dependence of $f_k$ will shortly be checked, but it
naturally follows from the fact that the dominant interactions of
gluons having momentum $k$ are with gluons having momentum on the order of $m$ and
with momentum transfer of the order of $m$ which is much less than
$k$. This means the Boltzmann equation will obey
\begin{equation}
{\partial \over \partial t} f_k(t) \sim m^3 \nabla_k^2 f_k(t)
\end{equation}
whose steady state solution is of the form (\ref{eq:f_k_1/k}). If we suppose that
$c(t)$ has a power dependence on $m_0 t$,
\begin{equation}
c(t) \sim \alpha^a (m_0 t)^b,
\label{eq:c_ans}
\end{equation}
then requiring $f_k(t) \sim 1/\alpha$ when $m_0 t \sim 1$ with $k\sim
m$ and requiring $f_k(t) \sim 1/\sqrt{\alpha}$ when $m_0 t \sim
\alpha^{-7/5}$ and $k \sim m$, as given by (\ref{eq:T_eq}) and (\ref{eq:f_k_app_1/k_wlk}), gives $a=0$
and $b=-5/14$ so that
\begin{equation}
c(t) \sim (m_0 t)^{-5/14} \ .
\label{eq:c_sol}
\end{equation}
Eqs.~(\ref{eq:m_fst}), (\ref{eq:f_k_1/k}) and (\ref{eq:c_ans}) give
\begin{equation}
{m\over p_0} \sim c(t) \sim (m_0t)^{-5/14}
\end{equation}
while (\ref{eq:p0_fst}) and (\ref{eq:c_sol}) give
\begin{equation}
m \sim m_0(m_0t)^{1/14} \ , 
\label{eq:m_is_exp}
\end{equation}
\begin{displaymath}
\quad p_0 \sim m_0 (m_0t)^{3/7} \ .
\end{displaymath}
It is now easy to check that the mean free path, $\lambda_{p_0}$, for gluons having mommentum $p_0$ obeys
\begin{equation}
{\lambda_p\over t} \sim \left[t p_0^3 f_{p_0}^2
\alpha^2/p_0^2\right]^{-1} = {p_0^3\over m^4t} \sim 1
\end{equation}
as one might expect. We can now see somewhat more sharply why
(\ref{eq:c_sol}) must hold. If $c(t)$ were to be parametrically larger
than that given by (\ref{eq:c_sol}) then $\lambda_{p_0}/t$ would be
less than one and the distribution (\ref{eq:f_k_1/k}) would not be
stable. If $c(t)$ were to be much smaller than given by
(\ref{eq:c_sol}) then the maximum value of $f_k$ which we would find,
for very small values of $k$, would significantly exceed that given in
(\ref{eq:fk3_is_res}) and would lead to the right hand side of
(\ref{eq:k3_fk3_1}) being much bigger than one, which is not
consistent.

Let us examine in a little more detail the consistency and stability
of (\ref{eq:f_k_1/k}).  In particular consider the emission or absorption of a gluon
$k,$ where $\bar{m}\ll k \ll p_0,$ by hard gluons as illustrated in
Fig.~\ref{inelas__23_32_scatt}. One has, as in (\ref{eq:2_3_in}),
\begin{equation}
\dot{f}_k \sim - {p_0^6\alpha^3\over m^3 k^3}
\left[f_{p_1}f_{p_2}f_k(1+f_{p_3})(1+f_{p_4})-f_{p_3}f_{p_4}
(1+f_{p_1})(1+f_{p_2})(1+f_k)\right].
\label{eq:fk_stab}
\end{equation}
Now the $f$'s are all large so we can limit ourselves to terms with
four factors of $f$ on the right-hand side of (\ref{eq:fk_stab}).  This yields
\begin{equation}
\dot{f}_k\sim - {p_0^6\alpha^3\over m^2 k^3} 
\{f_k\left[f_{p_1}f_{p_2}(f_{p_3}+f_{p_4})-f_{p_3}f_{p_4}(f_{p_1}+f_{p_2})\right]-
f_{p_1}f_{p_2}f_{p_3}f_{p_4}\}
\end{equation} 
Write
\begin{displaymath}
E_3 = E_1 + \omega \lambda
\end{displaymath}
\begin{equation}
E_4 = E_2 + \omega (1-\lambda)
\end{equation}
so that
\begin{equation}
\dot{f}_k \sim {p_0^6\alpha^3\over m^2 k^3} 
\{ (1-\lambda) f_{p_1}^2(f_{p_2}^{\prime}\omega f_k+f_{p_2}^2)+
\lambda f_{p_2}^2(f_{p_1}^{\prime}\omega f_k+ f_{p_1}^2)\}
\label{eq:fk_stab_der}
\end{equation}
where $f_{p_1}^{\prime}={\partial\over \partial E_1} f_{p_1},$ etc.
In (\ref{eq:fk_stab}) we have done a dimensional analysis.  The $p_0^6$ factor is a
rough way of writing the phase space factors coming from, say,
$d^3p_1d^3p_2.$ Thus the $f_{p_1}^2 (f_{p_2}^{\prime}\omega f_k +
f_{p_2}^2)$ in (\ref{eq:fk_stab_der}) should be understood as averaged over $p_1$ and
$p_2.$ Since $p_1$ and $p_2$ come in symmetrically we can rewrite (\ref{eq:fk_stab_der})
more simply as
\begin{equation}
\dot{f}_k\sim{p_0^6\alpha^3\over m^2 k^3} 
f_{p_1}^2 (f_{p_2}^{\prime}\omega f_k+ f_{p_2}^2)
\label{eq:fk_stab_sym}
\end{equation}
where $p_1,p_2$ are on the order of $p_0.$ Now the two terms,
 $f_p^{\prime}\omega f_k$ and $f_p^2,$ are of the same size if $f_k$
 has a $1/k$ behavior as indicated in (\ref{eq:f_k_1/k}).  Other power
 laws for $f_k$ will not allow $\dot{f}_k$ in (\ref{eq:fk_stab_sym})
 to be zero.  To see the stability of $\dot{f}_k=0$ suppose
 $f_{p_2}^{\prime}\omega f_k+f_{p_2}^2$ were positive.  Then $f_k$
 would grow in time.  But this would increase $f_{p_2}^{\prime}\omega
 f_k$ until it cancels $f_{p_2}^2.$ Thus (\ref{eq:f_k_1/k}) is a
 stable solution with respect to emission and absorption of gluons of
 momentum $k$ by harder gluons.  The exact value of $c,$ in
 (\ref{eq:f_k_1/k}), is determined by the vanishing of
 $<f_p^{\prime}\omega f_k+ f_p^2>.$

Now let's investigate the stability with respect to elastic scattering
$k_1+p_1\to k_2+p_2$ as illustrated in Fig.~\ref{elas_scatt}, and where $\bar{m}\ll
k_1, k_2\ll p_0.$ One has
\begin{equation}
\dot{f}_{k_2}^{1\leftrightarrow 2}\sim {p_0^3\alpha^2\over m^2}\\
\left[f_{p_1}f_{k_1}(1+f_{p_2})(1+f_{k_2}) - f_{p_2}f_{k_2}(1+f_{p_1})(1+f_{k_1})\right]
\label{eq:fk2_1to2_stab}
\end{equation}
where we include \underline{dimensional factors} for the $p_i$ and
$k_1$ phase space integrations, although we keep these momenta fixed
in the Bose-factors to better understand the stability of the solution
(\ref{eq:f_k_1/k}).  Calling $\omega_{k_1}-\omega_{k_2}=\Delta\omega$,
using (\ref{eq:f_k_1/k}) and writing
\begin{equation}
f_{p_1}-f_{p_2} \simeq - f_{p_1}^{\prime}\Delta\omega
\end{equation}
\begin{equation}
f_{k_2}-f_{k_1}\simeq {f_{k_1}f_{k_2}\over \omega f_\omega}\Delta\omega
\end{equation}
one can write (\ref{eq:fk2_1to2_stab}) as
\begin{equation}
\dot{f}_{k_2}^{1\leftrightarrow 2}\sim {-p_0^3\alpha^2\over m^2}{f_{k_1}f_{k_2}\Delta\omega\over \omega f_\omega}\left[f_{p_1}^{\prime}\omega f_\omega + f_{p_1}^2\right].
\end{equation}
Thus $\dot{f}_{k_2}^{1\leftrightarrow 2}$ will be zero under exactly
the same conditions on $c(t)$, in (\ref{eq:f_k_1/k}), as we found for stability
under emission and absorption as given in (\ref{eq:fk_stab_sym}).

Finally, we turn to the region $m<\omega < \bar{m}.$ The situation
here is very much like previously considered in Sec.~2.2.  The rate at
which gluons enter the system and then cascade down to a momentum
$k_3$ is given as in (\ref{eq:f_source})
\begin{equation}
\dot{f}_{k_3}^{source}\sim  {m\over \alpha}\left({m_0\over m}\right)^5 
\left({m\over k_3}\right)^3
\label{eq:f_k3_so}
\end{equation}
where now $m$ is given by (\ref{eq:m_is_exp}).  Eq.(\ref{eq:f_abs_k3_app}) is now changed to
\begin{equation}
\dot{f}_{k_3}^{abs} \sim - m{\sqrt{\alpha}} f_{k_3} \left[{1\over {\sqrt{\alpha}}} {m\over p_0}\right]
\label{eq:f_k3_abs}
\end{equation}
which is easily obtained from (\ref{eq:fk_stab_sym}) by dropping $f_{p_2}^2$ compared to
$f_{p_2}^{\prime}\omega f_k.$ Finally (\ref{eq:f_k3_3_to_4}) remains unchanged
\begin{equation}
\dot{f}_{k_3}^{3\rightarrow 4}\sim - m {\sqrt{\alpha}} f_{k_3}{\sqrt{\alpha}} f_{k_4}\left({k_4\over m}\right)^3 \left({k_3\over m}\right),
\label{eq:f_k3_34}
\end{equation}
and where the explanation of the meaning of this term is exactly as in
the previous section. Requiring that the $\dot{f}_{k_3}$ terms in
(\ref{eq:f_k3_so}), (\ref{eq:f_k3_abs}) and (\ref{eq:f_k3_34}) be of the same size gives
\begin{equation}
k_3/m \sim \left({m\over p_0}\right)^2 \left({m\over m_0}\right)^5 
\sim (m_0t)^{-5/14}
\label{eq:k3_514}
\end{equation}
and
\begin{equation}
f_{k_3}\sim {1\over \alpha}\left({p_0\over m}\right)^4 
\left({m_0\over m}\right)^{5}\sim {1\over \alpha}(m_0t)^{15/14}
\label{eq:fk3_is_res}
\end{equation}
which now replace (\ref{eq:k3_is}) and (\ref{eq:fk3_is}).  We note that (\ref{eq:k3_514}) and (\ref{eq:fk3_is_res}) agree
with (\ref{eq:k3_is}) and (\ref{eq:fk3_is}) when $m_0t=\alpha^{-7/5}.$
\begin{figure}[htb]
\setlength{\unitlength}{1.cm}
\begin{center}
\epsfig{file=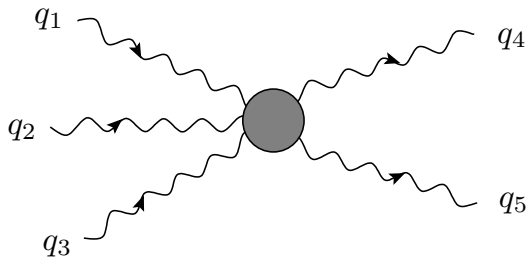, width=7cm}
\end{center}
\caption{Inelastic $3 \leftrightarrow 2$ gluon interactions.}
\label{inelas_32_scatt}
\end{figure}

Finally, we note that the process where three soft gluons, having
momenta on the order of $k_3$ as given in (\ref{eq:k3_514}),
annihilate into two harder gluons as illustrated in Fig.~\ref{inelas_32_scatt} with
$q_1,q_2, q_3\sim k_3$ gives an $\dot{f}_{k_3}$ of the same order as
(\ref{eq:f_k3_so}), (\ref{eq:f_k3_abs}) and (\ref{eq:f_k3_34}).
Indeed higher order processes where $n$ gluons having momenta on the
order of $k_3$ annihilate into 2 harder gluons also are of the same
order (see Appendix~\ref{App:inelastic_soft}) since 
\begin{equation}
\alpha\left ({k_3\over m}\right)^3 f_{k_3} \sim 1.
\label{eq:k3_fk3_1}
\end{equation}
%

\section{Fixing the source in terms of wavelengths}

In the discussion of the previous section we have assumed that the
source of incoming energy is spread uniformly in phase space in the
frequency range $m < \omega < \bar{m}$ where $m$ and $\bar{m}$ are of
the same order of magnitude.  Now we are going to express the source
in terms of wavelength rather than frequency.  We again suppose that
the rate of inflow of energy is
\begin{equation}
\dot{\epsilon} = m_0^5/\alpha,
\end{equation}
but we here suppose that the gluons are uniformly deposited in phase
space over the region
\begin{equation}
0 < k < k_0.
\label{eq:0_k_k0}
\end{equation}
That is we take $\dot{f}^{source}$ to be a constant, in both $\vec{k}$
and $t,$ in the region (\ref{eq:0_k_k0}) and where
\begin{equation}
{2(N_c^2-1)\over (2\pi)^3} \dot{f}^{source}\int_0^{k_0} 4\pi k^2\omega_k dk 
= m_0^5/\alpha \ .
\label{eq:f_dot_source}
\end{equation}
In the present situation we have two parameters $m_0$ and $k_0$ with
which to specify the source.  (By allowing $k_0$ to be time-dependent
the current specification includes the source used in Sec.~2.)  We
shall, however, always suppose that $k_0/m_0 \gtrsim 1$ as the
situation where $k_0$ is less than $m_0$ seems to have an abnormally
high rate of deposition of energy over a limited region of phase
space.  In the discussion that follows we deal with $m_0$ and $k_0$ as
time-independent quantities, however, it will be easy to allow a slow
time dependence of these quantities in our final results.

Much of the discussion of the last section is unchanged since the
basic pattern of the three different time regimes is set by the basic
structure of the problem.  We begin with the early time regime where
$m_0t < (1/\alpha)^{7/5}$.  From (\ref{eq:m_is_exp}) and (\ref{eq:k3_514}) we see that
\begin{equation}
{k_3\over k_0} \lesssim (m_0t)^{-2/7}
\end{equation}
and thus that $k_3/k_0 \ll 1.$ Thus the picture here is essentially
that described in Sec.~2.3, except that, when $k_0/m \ll 1,$ the region
$k_0 < k \lesssim m$ now has no incoming gluons from the
source. However gluons can come into this region from the process of
four gluons, having momentum on the order of $k_3$, going into a
gluon, $k$, and two harder gluons.  (We note that the three soft gluon
to two harder gluon process cannot give a gluon, $k$, in the region
$k_0 < k \ll m.$)

We can evaluate more precisely the dependence of $f_k,$ when $k_0<
k\ll m,$ by requiring a cancellation in the rate of gluons of momentum
$k$ produced by the competing processes.  First, the rate of
production of gluons coming from $q_1 + q_2 + q_3 + q_4 \rightarrow k
+ k_1 + k_2,$ with $q_1, q_2, q_3,$ and $q_4$ of order $k_3$ while
$k_1$ and $k_2$ are of order $m,$ is (see Appendix \ref{App:inelastic_soft})
\begin{equation}
\dot{f}_k^{4\rightarrow 3}\sim m f_k \alpha f_m
\label{eq:fk_4to3}
\end{equation}
where $\alpha\left({k_3\over m}\right)^3 f_{k_3}\sim 1$ has been used.
Secondly the rate of gluons absorbed by hard gluons is, from (\ref{eq:fk_stab_sym}), and
after letting the $1/k^3$ there become $1/m^3$ reflecting the cutoff
of the $1/k^3$ singularity when $k/m \ll 1$,
\begin{equation}
\dot{f}_k^{abs}\sim m f_k {c^2\over \alpha} 
\left(1/f_k - {\omega\,\alpha\over m\,c}\right)
\label{eq:f_abs_k_stab}
\end{equation}
where we have used
\begin{equation}
<f_p^{\prime}c(t)m/\alpha + f_p^2> = 0
\label{eq:stab_cond}
\end{equation}
the stablity condition determining $c$ that we found in Sec.2.3.  In
arriving at (\ref{eq:f_abs_k_stab}) we have also used $\left({p_0\over m}\right)^2\alpha
f_{p_0}\sim 1$ and $c=m/p_0$. Finally, there is the rate at which gluons
$k$ convert to gluons of momentum around $k_3$ by elastic scattering
with hard particles,
\begin{equation}
\dot{f}_k^{k\leftrightarrow k_3}\sim - p_0^3{\alpha^2\over m^2}
\left({k_3\over m}\right)^3{m\over k}\left[f_{p_1}f_k(1+f_{k_3})
(1+f_{p_2})-f_{k_3}f_{p_2}(1+f_k)(1+f_{p_1})\right].
\end{equation}
It is straightforward to get
\begin{equation}
\dot{f}_k^{k\leftrightarrow k_3}\sim p_0^3{\alpha\over m k}
[f_k(\omega-m)f_p^{\prime}+ f_p^2]
\end{equation}
which, after using (\ref{eq:stab_cond}), becomes
\begin{equation}
\dot{f}_k^{k \leftrightarrow k_3} = -m f_k {m\over k}\, 
{c\over \alpha}\left[{\alpha(\omega-m)\over m\,  c}-1/f_k\right].
\label{eq:fk_ktok3_stab}
\end{equation}
If $f_k$ were near ${c\,m\over \alpha\, \omega},$ its normal size as
given by (\ref{eq:f_k_1/k}), then (\ref{eq:fk_ktok3_stab}) would
dominate.  (Recall $c/{\sqrt{\alpha}}\,\sim
\left[\alpha^{7/5}m_0t\right]^{-5/14}$ is a large quantity in the time
domain we are currrently considering.)  This means that $1/f_k$ must
be much smaller than ``normal'', and the $1/f_k$ term can be dropped
in (\ref{eq:f_abs_k_stab}).  The remaining term in
(\ref{eq:f_abs_k_stab}) is comparable to (\ref{eq:fk_4to3}) so that one can get
stability by requiring 
\begin{displaymath}
-\omega \alpha/m \sim {m\over k}\left[{\alpha(\omega - m)\over m\, c}-1/f_k\right]
\end{displaymath}
or
\begin{equation}
f_k\sim {m  c\over \alpha}\ {1\over \omega - m + O([m_0t]^{-5/14}k)}
\end{equation}
so that $f_k$ begins to rise rapidly as $k$ decreases beyond  $m.$

Now turn to the region $m_0t > (1/\alpha)^{7/5}$ where gluons having
$\omega/m \gg 1$ are in thermal equilibrium. From (\ref{eq:k3_is})
\begin{equation}
k_3/k_0 \sim \left[\alpha^{9/5}m_0t\right]^{5/4} m/k_0 \ .
\end{equation}
So long as $k_3/k_0 \ll 1,$ that is so long as $m_0t\ll
(1/\alpha)^{9/5}(k_0/m)^{4/5},$ the large peak in $f_k$ will still
occur when $k\sim k_3$ and (\ref{eq:k3_is}) and (\ref{eq:fk3_is})
should remain valid.  Let us again investigate $f_k$ in the region
$k_0 < k \lesssim m$ when $k_3/k_0\ll 1.$ Now the three relevant
processes are
\begin{equation}
\dot{f}_k^{4\leftrightarrow 3}\sim m f_k\alpha f_m\left[\alpha f_3 
\left({k_3\over m}\right)^3\right]^4\sim m  f_k\alpha f_m 
\left[\alpha^{7/5}m_0t\right]^{-5}
\end{equation}
while
\begin{equation}
\dot{f}_k^{k\leftrightarrow k_3}\sim - {m^2\over k} f_k 
\left[\alpha^{7/5}m_0t\right]^{-5/4}\left[{\omega-m\over T}-1/f_k\right]T/m
\label{eq:fk_kto_k3_stab_fin}
\end{equation}
and the process where $k$ is emitted or absorbed by thermal particle
as illustrated in Fig.~\ref{inelas__23_32_scatt},
\begin{equation}
\dot{f}_k^{abs}\sim - m f_k{\sqrt{\alpha}}\left[{\omega\over T} - 1/f_k\right] T/m \ .
\label{eq:fk_dot_abs_stab_fin}
\end{equation}
The dominant contributions now are (\ref{eq:fk_kto_k3_stab_fin}) and
(\ref{eq:fk_dot_abs_stab_fin}).  Requiring a cancellation to get a
steady state gives
\begin{equation}
f_k\sim {T\over m}\left[(\omega-m)/m + O (k/m[\alpha^{9/5}m_0t]^{5/4})\right]^{-1}.
\label{eq:fk_eq_LO}
\end{equation}
Thus $f_k$ grows as $k$ decreases below $m$.  When $m_0t$ approaches
$(1/\alpha)^{9/5}(k_0/m)^{4/5}$ the picture starts to change as $k_3$
approaches $k_0.$ At $m_0t = (1/\alpha)^{9/5}(k_0/m)^{4/5}$ the
occupancy $ f_k$ begins to have little $k-$ dependence for $k/k_0 <
1.$ Requiring $\dot{f}_k^{source},$ as given by (\ref{eq:f_dot_source}), to cancel with
$\dot{f}_k^{abs},$ as given by (\ref{eq:fk_dot_abs_stab_fin}), in the domain $0 < k < k_0$ gives
\begin{equation}
{\sqrt{\alpha}} f_k \sim\left[\alpha^{9/5}m_0t\right]^{-5/4}(m/k_0)^3 \ .
\label{eq:fk_0_k_k0}
\end{equation}
We note that (\ref{eq:fk_eq_LO}), evaluated at $k=k_0$ is much smaller
than (\ref{eq:fk_0_k_k0}).  Thus for $m_0t >
(1/\alpha)^{9/5}(k_0/m)^{4/5}$ there is a ``discontinuity'' in $f_k$
with
\begin{equation}
{\sqrt{\alpha}} f_k\vert_{k=k_0-\epsilon}-{\sqrt{\alpha}} 
f_k\vert_{k=k_0+\epsilon}\sim\left[\alpha^{9/5}m_0t\right]^{-5/4}\left(m/k_0\right)^3 \ .
\end{equation}
By the time $m_0t \gg \alpha^{-9/5}(m/k_0)^{12/5}$ This jump in $f_k$
at $k_0$ has become very small, compared to the size of $f_k$, and
complete ``equilibration'' is reached.  Thus if $k_0$ is of the order
of $m$ then complete thermalization occurs at a time $m_0t \sim
(1/\alpha)^{9/5}$ as in Sec. 2.  If $k_0/m\ll 1$ then complete
thermalization occurs when
\begin{equation}
m_0t\sim(1/\alpha)^{9/5}(m/k_0)^{12/5} \ .
\end{equation}
%
%
\section*{Acknowledgements}

We would like to thank Stephane Munier for crucial suggestions and conversations
in the early stages of this work. A.~M. wishes to thank the DESY theory group
and the II Institute f\"ur Theoretische Physik of the University of Hamburg for
their hospitality while this work was being completed. He also wishes to thank
the Helmholtz and Humboldt foundations for a Helmholtz-Humboldt Research Award.
A.~Sh. also wishes to thank the DESY theory group for their hospitality during
his visit when this work was being finalized. A.~Sh. acknowledges financial
support by the Deutsche Forschungsgemeinschaft under contract Sh 92/2-1.

\appendix

\section{The derivation of Eq.~(\ref{eq:2_3_in})}
\label{App:2_3_in}
In this Appendix we consider the $2 \leftrightarrow 3$ gluon
interaction where $4$ gluons have momenta on the order of $T$ and the
momentum of the remaining gluon is $k \ll T$. To derive the result
given in Eq.~(\ref{eq:2_3_in}), we follow Ref.~\cite{Wong:2004ik}
where the $2 \leftrightarrow 3$ process was calculated in all
details. The most relevant process is the large angle gluon-gluon
scattering with a small angle gluon radiation/absorbtion (gluon-gluon
scattering with collinear emission/absorbtion). The inelastic $2
\leftrightarrow 3$ process is given by
\begin{eqnarray}
\dot{N} &=& 2\ \frac{[2(N_c^2-1)]^2}{(2\pi)^{15}}\ \int_R\ 
   {d^3p_1\over 2E_1}\ {d^3p_2\over 2E_2}\  
   {d^3p_3\over 2E_3}\ {d^3p_4\over 2E_4}\ {d^3k\over 2E_k}   
                \ (2\pi)^4\delta^4(p_1+p_2-p_3-p_4-k)
\nonumber \\
&& \times 
                \vert M_{12\to34k}\vert^2 
                \ \left[f_{p_1}f_{p_2}f_{k}(1+f_{p_3})(1+f_{p_4})-
                f_{p_3}f_{p_4}(1+f_{p_1})(1+f_{p_2})(1+f_{k})\right]
\nonumber\\
\label{App:N_2to3}
\end{eqnarray}
where $\dot{N} \equiv 2 (N^2_c-1) \int \frac{d^3k}{(2\pi)^3} \dot{f}_k$ and the
symbol $R$ in the integration in (\ref{App:N_2to3}) indicates that $k
\ll p_1, p_2, p_3, p_4$. To evaluate (\ref{App:N_2to3}) one supposes
that the $5$-gluon process consists of a hard collision involving four
gluons. Then the fifth gluon, $k$, is collinear with one of the
incoming or outgoing gluons, giving a factor of $4$ which when devided
by the symmetry factor, $2$, explains the appearance of $2$ in
(\ref{App:N_2to3}).  Since all four possibilities lead to the same
result, let us consider only the case where the gluons $4$ and $k$ are
collinear. Collinear means that the invariant mass $s_{4k} =(p_4+k)^2$
is significantly less than $Q^2 = \mbox{min}(|s|, |t|, |u|)$, where
$Q$ is the momentum scale of the hard scattering.

In Ref.~\cite{Wong:2004ik} (see Eq.~(20)) it has been shown that the
following factorisation holds in the collinear limit 
\begin{eqnarray}
\!\!&&\int \frac{d^3p_3}{(2\pi)^3 E_3}\ \frac{d^3p_4}{(2\pi)^3 E_4}\ 
\frac{d^3k}{(2\pi)^3 E_k}\ (2\pi)^4 \delta^4(p_1+p_2-p_3-p_4-k)\ \vert M_{12\to34k}\vert^2
\nonumber \\
\!\!\!\simeq\!\!\!\!\!\!\!\!\!&&\int\!\frac{d^3p_3}{(2\pi)^3 E_3}\,
\frac{d^3p_4}{(2\pi)^3 E_4}\,
(2\pi)^4 \delta^4(p_1\!+\!p_2\!-\!p_3\!-\!p_4)\,\vert M_{12\to34}\vert^2\!\!\times\!\! 
\int\!\frac{dzds_{4k}}{s_{4k}}\,\frac{\alpha_s(s_{4k})}{2\pi}\,P_{gg}(z) 
\nonumber \\
\label{App:stephen}
\end{eqnarray}
where $P_{gg}(z)$ is the unregularized $g \to gg$ splitting function
\begin{equation}
P_{gg}(z) = 2 N_c \left( \frac{1-z}{z} + \frac{z}{1-z} + z(1-z)\right)
\label{App:P_gg}
\end{equation}
and $\vert M_{12\to34}\vert^2$ is the matrix element for the hard
scattering given by (\ref{eq:M}). With (\ref{App:stephen}) Eq.~(\ref{App:N_2to3}) becomes
\begin{eqnarray}
&&\!\!\!\!\!\!\!\!\!\dot{N} = 2\ \frac{[2(N_c^2-1)]^2}{(2\pi)^{13}}\ \int_R\ 
   {d^3p_1\over 2E_1}\ {d^3p_2\over 2E_2}\  
   {d^3p_3\over 2E_3}\ {d^3p_4\over 2E_4}\   
                \ (2\pi)^4 \delta^4(p_1+p_2-p_3-p_4)\ \vert M_{12\to34}\vert^2\
\nonumber \\
&&\!\!\!\!\!\!\!\!\!\times\! 
                \int_R\!\frac{dzds_{4k}}{s_{4k}}\,\frac{\alpha_s(s_{4k})}{2\pi}\,P_{gg}(z)
                \left[f_{p_1}f_{p_2}f_{k}(1\!+\!f_{p_3})(1\!+\!f_{p_4})-
                f_{p_3}f_{p_4}(1\!+\!f_{p_1})(1\!+\!f_{p_2})(1\!+\!f_{k})\right].
\nonumber\\
\label{App:N_2to3_eval}
\end{eqnarray}
The dominant contribution to $|M_{12\to34}|$ comes from the small-$t$ region
\begin{equation}
\vert M\vert^2 = \frac{64\pi^4}{(N_c^2-1)}\left({\alpha N_c\over \pi}\right)^2 \ 
\left[-\frac{us}{t}\right] \ ,
\end{equation}
where $u \approx -s$ for small $t=(p_1-p_3)^2$ due to $s+u+t=0$. Using also 
$s=(p_1+p_2)^2 \approx 2 E^2$, the integration over $p_4$ in (\ref{App:N_2to3_eval}) 
gives
\begin{eqnarray}
&&\!\!\!\!\!\!\!\!\!\dot{N} = \frac{[2(N_c^2-1)]^2}{4(2\pi)^{5}}\ 
   \left({\alpha N_c\over \pi}\right)^2 \int_R\ 
   {d^3p_1}\ {d^3p_2}\  
   {d^3p_3}\ \delta^4(E_1+E_2-E_3-E_4)\ \frac{1}{[(p_1-p_3)^2]^2} 
\nonumber \\
&&\!\!\!\!\!\!\!\!\!\times\! 
                \int_R\!\frac{dzds_{4k}}{s_{4k}}\,\frac{\alpha_s(s_{4k})}{2\pi}\,P_{gg}(z)
                \left[f_{p_1}f_{p_2}f_{k}(1\!+\!f_{p_3})(1\!+\!f_{p_4})-
                f_{p_3}f_{p_4}(1\!+\!f_{p_1})(1\!+\!f_{p_2})(1\!+\!f_{k})\right].
\nonumber\\
\label{App:N_2to3_eval_Meval}
\end{eqnarray}
With $d^3p_3 = d^2p_{3\perp}dp_{3L} = d^2p_{3\perp}dE_3$, where $p_{3\perp}$
is the transverse and $p_{3L}$ the longintudinal component of $p_3$,
and $(p_1-p_3)^2=p_{3\perp}^2$, one can perform also the integration
over $dE_3$ and obtains  
\begin{eqnarray}
&&\!\!\!\!\!\!\!\!\!\dot{N} = \frac{[2(N_c^2-1)]^2}{4(2\pi)^{5}}\ 
   \left({\alpha N_c\over \pi}\right)^2 \int_R\ 
   {d^3p_1}\ {d^3p_2}\ {d^2p_{3\perp}}\ \frac{1}{p_{3\perp}^4} 
\nonumber \\
&&\!\!\!\!\!\!\!\!\!\times\! 
                \int_R\!\frac{dzds_{4k}}{s_{4k}}\,\frac{\alpha_s(s_{4k})}{2\pi}\,P_{gg}(z)
                \left[f_{p_1}f_{p_2}f_{k}(1\!+\!f_{p_3})(1\!+\!f_{p_4})-
                f_{p_3}f_{p_4}(1\!+\!f_{p_1})(1\!+\!f_{p_2})(1\!+\!f_{k})\right].
\nonumber\\
\label{App:N_2to3_eval_Meval_de}
\end{eqnarray}
Now, with $p_{3\perp}^2 \sim m^2$ because of screening, $d^2p_{3\perp}
\sim m^2$, $d^3p_1 \sim T^3$, $d^3p_2 \sim T^3$ and the relation
$\dot{\epsilon} = \dot{N} w$, one obtains the parametrical estimate
given in Eq.~(\ref{eq:2_3_in}). Note that $k \ll p_1, p_2,
p_3, p_4$ is satisfied by requiring $z$ to take only values close to
$0$ in (\ref{App:N_2to3_eval_Meval_de}).

\section{The derivation of Eqs.~(\ref{eq:3_4_el}), (\ref{eq:fk5_5_3}) and 
(\ref{eq:fk3_1_3})}
\label{App:3_4_el}
In this Appendix we show how to get the parametrical estimates given
in Eqs.~(\ref{eq:3_4_el}), (\ref{eq:fk5_5_3}) and
(\ref{eq:fk3_1_3}). Let us first derive Eq.~(\ref{eq:3_4_el}). The
elastic scattering of soft gluons, $k_3$ and $k_4$ ($k_3>k_4$), with
hard gluons having momenta $p_1$, $p_2$ on the order of $T$, as shown
in Fig.~\ref{elas_scatt} with the replacement $q_1 \to k_3$ and $q_2
\to k_4$, is described by 
\begin{eqnarray}
\dot{f}_{k_3}^{3\to 4} &=&-\frac{2 (N_c^2-1)}{(2\pi)^5\, 2\omega_3}
                       \int {d^3p_1\over 2E_1} \ 
                 {d^3p_2\over 2E_2}\ {d^3k_4\over 2\omega_4} 
                 \ \delta^4(p_1+k_3-p_2-k_4)\ \vert M\vert^2  
\nonumber \\
&& \hspace*{0.6cm} \times 
                \left[f_{k_3}f_{p_1}(1+f_{k_4})(1+f_{p_2})-
                f_{k_4}f_{p_2}(1+f_{k_3})(1+f_{p_1})\right] \ .
\label{eq:f_3_4}
\end{eqnarray}
where $|M|^2$ is given in Eq.~(\ref{eq:M}). The hard momenta read $p_i
=(E_i,\vec{p}_i)$ with $E_i \approx |\vec{p}_i|$ while the soft momenta are
$k_j=(\omega_j,\vec{k}_j)$ with $\omega_j^2 = m^2+\vec{k}_j^2 \approx
m^2$.

It is the small angle scattering ($t$ small) which gives the dominant
contribution to $|M|^2$,
\begin{equation}
\vert M\vert^2 = \frac{64\pi^4}{(N_c^2-1)}\,\left({\alpha N_c\over \pi}\right)^2 \ 
\left[-\frac{us}{t}\right] \ .
\end{equation}
Taking into account that $t \sim m^2$ because of screening, $u \approx
-s$ due to $s+u+t=0$, and $s=(p_1+k_3)^2 \approx 2 E_1 \omega_3$, one
obtains
\begin{equation}
\vert M\vert^2 \approx \frac{64\pi^4}{(N_c^2-1)}\,\left({\alpha N_c\over \pi}\right)^2 \ \frac{(2\,E\,m)^2}{m^4} \ .
\end{equation}
Using also  
\begin{equation}
\delta^4(p_1+k_3-p_2-k_4) = \delta^3(\vec{p}_1+\vec{k}_3-\vec{p}_2-\vec{k}_4) 
                             \ \delta(E_1+\omega_3-E_2-\omega_4) 
\end{equation}
the integration over $p_2$ in Eq.~(\ref{eq:f_3_4}) then gives
\begin{eqnarray}
\dot{f}_{k_3}^{3\to 4} &\approx&-\frac{16}{(2\pi)^3}\frac{(\alpha N_c)^2}{m^4}
                       \int {d^3p_1} \ {d^3k_4} 
 \ \delta(E_1+\omega_3-\sqrt{(\vec{p}_1+\vec{k}_3-\vec{k}_4)^2}-\omega_4)   
\nonumber \\
&& \hspace*{0.6cm} \times 
                \left[f_{k_3}f_{p_1}(1+f_{k_4})(1+f_{p_2})-
                f_{k_4}f_{p_2}(1+f_{k_3})(1+f_{p_1})\right] \ ,
\label{eq:f_3_4_app}
\end{eqnarray}
where $f_{p_2}$ within the brackets $[\ ]$ is to be evaluated at ${\vec{p}_2=\vec{p}_1+\vec{k}_3-\vec{k}_4}$. 

The evaluation of the delta-function  
\begin{eqnarray}
&&\delta(E_1+\omega_3-\sqrt{(\vec{p}_1+\vec{k}_3-\vec{k}_4)^2}-\omega_4) 
\nonumber\\
&&\quad \approx  \delta(|\vec{p}_1| -m +\frac{\vec{k}^2_3}{2m} - |\vec{p}_1|-
\frac{\vec{p}_1}{|\vec{p}_1|} (\vec{k}_3-\vec{k}_4) -m - \frac{\vec{k}^2_4}{2m}) \nonumber \\
&&\quad \approx \delta(-|\vec{k}_3| \cos(\theta)) \ ,
\end{eqnarray}
where $\theta$ is the angle between $\vec{p}_1$ and $\vec{k}_3$,
allows us to perform also the angular integration, $d^3p_1
\delta(-|\vec{k}_3| \cos(\theta)) = 2\pi p_1^2
d|\vec{p}_1|/|\vec{k}_3|$, which leads to
\begin{eqnarray}
\dot{f}_{k_3}^{3\to 4} &\approx&- \frac{16}{(2\pi)^2}\frac{(\alpha N_c)^2}{m^4}
                       \int p_1^2 d|\vec{p}_1| \ {d^3k_4}\ {1 \over |\vec{k}_3|}   
\nonumber \\
&& \times 
                \left[f_{k_3}f_{p_1}(1+f_{k_4})(1+f_{p_2})-
                f_{k_4}f_{p_2}(1+f_{k_3})(1+f_{p_1})\right] \ .
\label{eq:f_3_4_app_del}
\end{eqnarray}
Now, with $p_1^2 \sim T^2$ and $d|\vec{p}_1| \sim T$, is easy to see
that Eq.~(\ref{eq:f_3_4_app_del}) is parametrically the same as
Eq.~(\ref{eq:3_4_el}).

To get Eq.~(\ref{eq:fk5_5_3}), one follows the same procedure as
above, with the replacement $k_3 \rightarrow k_5$ and $k_4 \rightarrow
k_3$, and gets
\begin{eqnarray}
\dot{f}_{k_5}^{5\to 3} &\approx&- \frac{16}{(2\pi)^2}\frac{(\alpha N_c)^2}{m^4}
                       \int p_1^2 d|\vec{p}_1| \ {d^3k_3}\ {1 \over |\vec{k}_5|}   
\nonumber \\
&& \times 
                \left[f_{k_5}f_{p_1}(1+f_{k_3})(1+f_{p_2})-
                f_{k_3}f_{p_2}(1+f_{k_5})(1+f_{p_1})\right] \ .
\label{eq:f_5_3_app_del}
\end{eqnarray}
Using (\ref{eq:fp_eq}) which allows us to write the bracket in
(\ref{eq:f_5_3_app_del}) as
\begin{equation}
\left[\ \right] 
= e^{E_{1}/T} f_{p_1} f_{p_2}f_{k_5}f_{k_3} \left[(\omega_5-\omega_3)/T+
1/f_{k_3}\ -1/f_{k_5}\right] \ 
\end{equation}
and $p_1, p_2 \sim T$, $d^3k_3 \sim k^3_3$ and (\ref{eq:freq}), we get  
\begin{equation}
\dot{f}_{k_5}^{5\rightarrow 3}\sim - m \sqrt{\alpha} 
f_{k_5} \left[\sqrt{\alpha} f_{k_3}\left({k_3\over m}\right)^3\right] \left(\frac{m}{k_5}\right) \ 
\left[{\omega_5-\omega_3\over T}+1/f_{k_3} -1/f_{k_5}\right] T/m 
\label{app:fk5_5_3_fin_es}
\end{equation}
which for $m/k_5 \sim 1$, $w_3 \approx m$ and $1/f_{k_3} \ll
1/f_{k_5}$ reduces to the result in Eq.~(\ref{eq:fk5_5_3}).

To get Eq.~(\ref{eq:fk3_1_3}), one starts with
\begin{eqnarray}
\dot{f}_{k_3}^{1\to 3} &=&\frac{2(N_c^2-1)}{(2\pi)^5\,2\omega_3}
                       \int {d^3p_1\over 2E_1} \ 
                 {d^3p_2\over 2E_2}\ {d^3k_1\over 2\omega_1} 
                 \ \delta^4(p_1+k_1-p_2-k_3)\ \vert M\vert^2  
\nonumber \\
&& \hspace*{0.6cm} \times 
                \left[f_{k_1}f_{p_1}(1+f_{k_3})(1+f_{p_2})-
                f_{k_3}f_{p_2}(1+f_{k_1})(1+f_{p_1})\right] 
\label{app:fk3_1_3}
\end{eqnarray}
and follows the same steps as in above calculations which lead to
\begin{equation}
\dot{f}_{k_3}^{1\rightarrow 3} \sim m {\sqrt{\alpha}} f_{k_1} 
{\sqrt{\alpha}} f_{k_3} \left(\frac{k_1}{m}\right)^2 \left[{\omega_1-m\over T}-1/f_{k_1}\right] T/m \ ,
\label{app:fk3_1_3_fin}
\end{equation}
where the factor $(k_1/m)^2$ comes from the phase space $d^3k_1$ and
the angular integration $d^3p_1\,\delta(-|\vec{k}_1|\cos(\theta)) =
2\pi\,p_2^2\,d|\vec{p}_2|/|\vec{k}_1|$. Since $k_1$ is close to $m$ in
the case discussed, it is legitimate to set $(k_1/m)^2 \sim 1$ which
gives the estimate in Eq.~(\ref{eq:fk3_1_3}).

\section{$n \to 2$ gluon merging}
\label{App:inelastic_soft}
Here we show that the process where $n$ soft gluons, having momenta on
the order of $k_3$ as given in Eq.~(\ref{eq:k3_514}), merge into two
harder gluons ( with $k_1$, $k_2 \sim m$) gives an $\dot{f}_{k_3}$ on
the same order as (\ref{eq:f_k3_so}), (\ref{eq:f_k3_abs}) and
(\ref{eq:f_k3_34}). It is instructive to begin with a simpler process
where only $3$ three soft gluons annihilate into two harder gluons as
shown in Fig.~\ref{inelas_32_scatt}, with $q_1,q_2, q_3\sim k_3$, and then go over to the more general case. For the $3 \rightarrow 2$ process we have   
\begin{eqnarray}
\dot{f}^{3\rightarrow 2}_{k_3} &=& \frac{2(N_c^2-1)}{(2\pi)^8\,2\omega_3} \int\ 
   {d^3q_1\over 2\omega_1}\ {d^3q_2\over 2\omega_2}\  
   {d^3k_1\over 2\omega_{k_1}}\ {d^3k_2\over 2\omega_{k_2}} 
                \ \delta^4(q_1+q_2+q_3-k_1-k_2)\ \vert M\vert^2
\nonumber \\
&& \times 
                \left[f_{q_1}f_{q_2}f_{q_3}(1+f_{k_1})(1+f_{k_2})-
                f_{k_1}f_{k_2}(1+f_{q_1})(1+f_{q_2})(1+f_{q_3})\right] 
\end{eqnarray}
and after the dimensional estimate 
\begin{equation}
|M^2| \sim \alpha^3 \frac{1}{m^2} \ 
\end{equation}
and the integration over $k_2$
\begin{eqnarray}
\dot{f}^{3\rightarrow 2}_{k_3} &\sim& \frac{1}{m^7} \int\ 
   {d^3q_1}\ {d^3q_2}\ {d^3k_1}\ 
\delta(\omega_{q_1}+\omega_{q_2}+\omega_{q_3}-\omega_{k_1}-\omega_{k_2})\ 
\nonumber \\
&& \times 
                \left[f_{q_1}f_{q_2}f_{q_3}(1+f_{k_1})(1+f_{k_2})-
                f_{k_1}f_{k_2}(1+f_{q_1})(1+f_{q_2})(1+f_{q_3})\right] \ .
\end{eqnarray}
Doing another dimensional estimate 
\begin{equation}
{d^3k_1}\ 
\delta(\omega_{q_1}+\omega_{q_2}+\omega_{q_3}-\omega_{k_1}-\omega_{k_2})
\sim m^2 ,
\end{equation}
keeping only the leading contribution in the brackets
\begin{eqnarray}
&&\left[f_{q_1}f_{q_2}f_{q_3}(1+f_{k_1})(1+f_{k_2})-
                f_{k_1}f_{k_2}(1+f_{q_1})(1+f_{q_2})(1+f_{q_3})\right] 
\nonumber \\
&& = \left[f_{k_3}^3 \ f_m\right] + \mbox{corrections}
\end{eqnarray}
and using $d^3q_1$, $d^3q_2 \sim k_3^3$, we find 
\begin{equation}
\dot{f}^{3\rightarrow 2}_{k_3} \sim m f_m \alpha f_k \left[\alpha
\left(\frac{k_3}{m}\right) f_{k_3}\right]^3 
\label{app:fk3_sfin}
\end{equation}
which with (\ref{eq:k3_fk3_1}) reduces to 
\begin{equation}
\dot{f}^{3\rightarrow 2}_{k_3} \sim m f_m \alpha f_k 
\label{app:fk3_fin}
\end{equation}
and is of the same order as (\ref{eq:f_k3_so}), (\ref{eq:f_k3_abs}) and
(\ref{eq:f_k3_34}).

Following the same procedure, it is easy to show for
the $n \rightarrow 2$ process that 
\begin{equation}
\dot{f}^{n\rightarrow 2}_{k_3} \sim m f_m \alpha f_k \left[\alpha
\left(\frac{k_3}{m}\right) f_{k_3}\right]^n 
\label{app:fk3_n_2_sfin}
\end{equation}
which again, when using (\ref{eq:k3_fk3_1}), becomes of the same order as 
(\ref{eq:f_k3_so}), (\ref{eq:f_k3_abs}) and (\ref{eq:f_k3_34}). 
%
  
%


\begin{thebibliography}{99}

\bibitem{Kolb:2001qz+X}
  P.~F.~Kolb, U.~W.~Heinz, P.~Huovinen, K.~J.~Eskola and K.~Tuominen,
  Nucl.\ Phys.\ A {\bf 696} (2001) 197;\ 
  P.~Huovinen, P.~F.~Kolb, U.~W.~Heinz, P.~V.~Ruuskanen and S.~A.~Voloshin,
  Phys.\ Lett.\ B {\bf 503} (2001) 58;\ 
  P.~F.~Kolb, P.~Huovinen, U.~W.~Heinz and H.~Heiselberg,
  Phys.\ Lett.\ B {\bf 500} (2001) 232;\ 
  P.~F.~Kolb, J.~Sollfrank and U.~W.~Heinz,
  Phys.\ Rev.\ C {\bf 62} (2000) 054909.

\bibitem{Teaney:2003kp+X}
  D.~Teaney,
  Phys.\ Rev.\ C {\bf 68} (2003) 034913;\ 
  D.~Teaney, J.~Lauret and E.~V.~Shuryak,
  arXiv:nucl-th/0110037; \ 
  Phys.\ Rev.\ Lett.\  {\bf 86} (2001) 4783.

\bibitem{Baier:2000sb}
  R.~Baier, A.~H.~Mueller, D.~Schiff and D.~T.~Son,
  Phys.\ Lett.\ B {\bf 502} (2001) 51.

\bibitem{Wong:1996va}
  S.~M.~H.~Wong,
  Phys.\ Rev.\ C {\bf 54} (1996) 2588.


\bibitem{Baier:2002bt}
  R.~Baier, A.~H.~Mueller, D.~Schiff and D.~T.~Son,
  Phys.\ Lett.\ B {\bf 539} (2002) 46.

\bibitem{Arnold:2003rq}
  P.~Arnold, J.~Lenaghan and G.~D.~Moore,
  JHEP {\bf 0308} (2003) 002.



\bibitem{Mrowczynski:1988dz+X}
  S.~Mrowczynski,
  Phys.\ Lett.\ B {\bf 214} (1988) 587;\ 
  Phys.\ Lett.\ B {\bf 314} (1993) 118:\ 
  Phys.\ Rev.\ C {\bf 49} (1994) 2191;\ 
  Phys.\ Lett.\ B {\bf 393} (1997) 26; \\ 
  S.~Mrowczynski and M.~H.~Thoma,
  Phys.\ Rev.\ D {\bf 62} (2000) 036011;\\
  J.~Randrup and S.~Mrowczynski,
  Phys.\ Rev.\ C {\bf 68} (2003) 034909.

\bibitem{Weibel:1959}
 E.~S.~Weibel,
  Phys. Rev. Lett. {\bf 2}, 83 (1959).

\bibitem{Buneman:1958}
 O.~Buneman,
  Phys. Rev. Lett. {\bf 1}, 8 (1958).

\bibitem{Heinz:1985vf+X}
  U.~W.~Heinz,
  Nucl.\ Phys.\ A {\bf 418} (1984) 603C;\\
  Y.~E.~Pokrovsky and A.~V.~Selikhov,
  JETP Lett.\  {\bf 47} (1988) 12
  [Pisma Zh.\ Eksp.\ Teor.\ Fiz.\  {\bf 47} (1988) 11];
  Sov.\ J.\ Nucl.\ Phys.\  {\bf 52} (1990) 146
  [Yad.\ Fiz.\  {\bf 52} (1990) 229];
  Sov.\ J.\ Nucl.\ Phys.\  {\bf 52} (1990) 385
  [Yad.\ Fiz.\  {\bf 52} (1990) 605];\\
  O.~P.~Pavlenko,
  Sov.\ J.\ Nucl.\ Phys.\  {\bf 55} (1992) 1243
  [Yad.\ Fiz.\  {\bf 55} (1992) 2239].

\bibitem{Mueller:2005un+X}
  A.~H.~Mueller, A.~I.~Shoshi and S.~M.~H.~Wong,
  Phys.\ Lett.\ B {\bf 632} (2006) 257;\ 
  ``A modified 'bottom-up' thermalization in heavy ion collisions,''
  arXiv:hep-ph/0512045.

\bibitem{Bodeker:2005nv}
  D.~Bodeker,
  JHEP {\bf 0510} (2005) 092.

\bibitem{Arnold:2005ef}
  P.~Arnold and G.~D.~Moore,
  Phys.\ Rev.\ D {\bf 73} (2006) 025006. 

\bibitem{Romatschke:2003ms+X}
  P.~Romatschke and M.~Strickland,
  Phys.\ Rev.\ D {\bf 68} (2003) 036004;\ 
  P.~Romatschke and M.~Strickland,
  Phys.\ Rev.\ D {\bf 70} (2004) 116006.

\bibitem{Rebhan:2004ur+X}
  A.~Rebhan, P.~Romatschke and M.~Strickland,
  Phys.\ Rev.\ Lett.\  {\bf 94} (2005) 102303;\ 

\bibitem{Dumitru:2005gp+X}
  A.~Dumitru and Y.~Nara,
  Phys.\ Lett.\ B {\bf 621} (2005) 89;\ 
  A.~Dumitru, Y.~Nara and M.~Strickland,
  ``Ultraviolet avalanche in anisotropic non-Abelian plasmas,''
  arXiv:hep-ph/0604149.

\bibitem{Romatschke:2005pm+X}
  P.~Romatschke and R.~Venugopalan,
  Phys.\ Rev.\ Lett.\  {\bf 96} (2006) 062302;\ 
  ``The unstable Glasma,''
  arXiv:hep-ph/0605045;\ 
  P.~Romatschke and A.~Rebhan,
  ``Plasma instabilities in an anisotropically expanding geometry,''
  arXiv:hep-ph/0605064.

\bibitem{Mrowczynski:2005ki}
  S.~Mrowczynski,
  %
  Acta Phys.\ Polon.\ B {\bf 37} (2006) 427. 

\bibitem{Arnold:2005qs}
  P.~Arnold and G.~D.~Moore,
  Phys.\ Rev.\ D {\bf 73} (2006) 025013.


\bibitem{ZLF}
V. Zakharov, V. L'vov, and G. Falkovich,
  ``Kolmogorov Spectra of Turbulence, Wave Turbulence'',
  Springer-Verlag, Berlin, 1992. 

\bibitem{Micha:2004bv}
  R.~Micha and I.~I.~Tkachev,
  Phys.\ Rev.\ D {\bf 70} (2004) 043538.


\bibitem{Wong:2004ik}
  S.~M.~H.~Wong,
  ``Out-of-equilibrium collinear enhanced equilibration in the bottom-up
  thermalization scenario in heavy ion collisions,''
  arXiv:hep-ph/0404222.



%
\end{thebibliography}
\end{document}